\documentclass[preprint,aps,pra]{revtex4}

\usepackage{graphicx}

\begin{document}

\title{Thermoelectric performance in electron and hole doped PtSb$_2$}

\author{Y. Saeed$^{1}$, N. Singh, Schwingenschl\"ogl$^{1}$}

\affiliation{$^{1}$Physical Science \& Engineering division, KAUST, Thuwal 23955-6900, Kingdom of Saudi Arabia}

\begin{abstract}
We employ density functional theory to investigate the thermoelectric properties
of electron and hole doped PtSb$_2$. Our results show that for doping of 0.04 holes per unit cell (1.5$\times10^{20}$ cm$^{-3}$) 
PtSb$_2$ shows a high Seebeck coefficient at room temperature, which can also be achieved 
at other temperatures by controlling the carrier concentration (both electron and hole). 
The electrical conductivity becomes temperature independent when the doping exceeds about 0.20 electrons/holes 
per unit cell. The figure of merit at 800 K in electron and hole doped PtSb$_2$ is 0.13 and 0.21, respectively. The thermoelectric efficiency with same host material are predicted for certain doping levels.   

\end{abstract}


\maketitle

\section{Introduction}

Efficient room-temperature thermoelectric materials are currently a striking challenge for researchers.
In general, the efficiency is determined by the figure of merit $ZT={\sigma}S^{2}T/\kappa$, 
where $\sigma$ is the electrical conductivity, $S$ is the Seebeck coefficient, $T$ is the temperature,
and $\kappa$ is the thermal conductivity. To achieve a high $ZT$, $\sigma$ and $S$ should be large while at the same 
time $\kappa$ should be small. However, increasing $\sigma$ by increasing the carrier concentration usually decreases $S$ and 
increases $\kappa$. Therefore, it is a prime task to control the numerator $\sigma S^2$ (the power factor) and the denominator $\kappa$ independently.

Kuroki $et$ $al.$ have shown that a metal with pudding mold type bands (a dispersive portion and a flat portion) 
can exhibit good thermoelectric properties \cite{Kuroki}. Recently, such bands have been reported in the 
cubic pyrite material PtSb$_2$, which has a high $S=250$ $\mu$VK$^{-1}$ \cite{Nishikubo}. Bulk PtSb$_2$ can be both metallic and semiconducting for deviations from the ideal stoichiometry and it can be \textit{n}- and \textit{p}-type doped with charge carrier concentrations spanning several orders of magnitude (10$^{16}$ cm$^{-3}$ to 10$^{20}$ cm$^{-3}$), as observed for different synthesis methods and parameters \cite{Damon, Elliott, Abdullaev, Dargys, Laudise, Nikolic}. While $p$-type PtSb$_2$ (by Ir doping on the Pt site) gives a high metallic conductivity and large Seebeck coefficient \cite{Nishikubo}, $n$-type PtSb$_2$ (by Sb deficiency) gives a large Seebeck coefficient and thermal conductivity between 0 and 300 K \cite{Sondergaard}. PtSb$_2$ has a high mobility due to the small difference in the electronegativities of Pt and Sb (2.28 and 2.05, respectively) \cite{Slack}.  

Thermoelectric device performance relies directly
on the temperature difference $\triangle{T}=T_{hot}-T_{cold}$ and the intrinsic material
parameter $ZT$. For a thermoelectric generator, the thermoelectric efficiency is defined by
the Carnot efficiency $\triangle{T}/T_{hot}$ and the figure of
merit as
\begin{equation}
\eta=\frac{\triangle{T}}{T_{hot}}\left(\frac{\sqrt{1+ZT_{avg}}-1}{\sqrt{1+ZT_{avg}}+\frac{T_{cold}}{T_{hot}}}\right)
\end{equation}
where $T_{hot}$ and $T_{cold}$ are the temperatures of the hot and cold ends of the device and $T_{avg}=(T_{hot}- T_{cold})/2$. The figure of merit of a thermoelectric device is
\begin{equation}
ZT_{avg}=\left(\frac{S_{h}-S_{e}}{\sqrt{\frac{\kappa_{h}}{\sigma_{h}}}+\sqrt{\frac{\kappa_{e}}{\sigma_{e}}}}\right)^{2}T_{avg} 
\end{equation}
where $h$ and $e$ represent hole and electron doping, respectively. 

The term $\sqrt{1+ZT_{avg}}$ varies with the average temperature, indicating that an increasing efficiency requires both a high $ZT_{avg}$ and a large $\triangle{T}$.
 Thermoelectric devices currently available reach $ZT_{avg}=0.8$ and operate at an efficiency of 5 to 6$\%$ \cite{Sootsman}. By increasing the figure of merit by a factor
 of 4 (depending on $\triangle$T) the predicted efficiency increases to 30$\%$, a highly attractive prospect. Different host materials have different work functions,
 $i.e.$, they have different potential barriers and resistances at the metal contact in the thermocouple, which affects the mobility of the carriers. Therefore, it is 
advantageous to use the same material for the electron and hole doped regions. In this respect, doped PtSb$_2$ is a promising thermoelectric material with bipolar conduction.
 In the present work, we address the effect of electron and hole doping in PtSb$_2,$ which has a melting temperature of 1500 K \cite{Damon2} on $\sigma$ and $S$ from 300 K to 800 K. We study the power factor and the figure of merit for both types of doping. We also calculate the thermoelectric generator efficiency 
by Eqs. (1) and (2). Our calculations demonstrate excellent materials properties for doped PtSb$_2$.

\begin{figure}
\includegraphics[width=0.5\textwidth,clip]{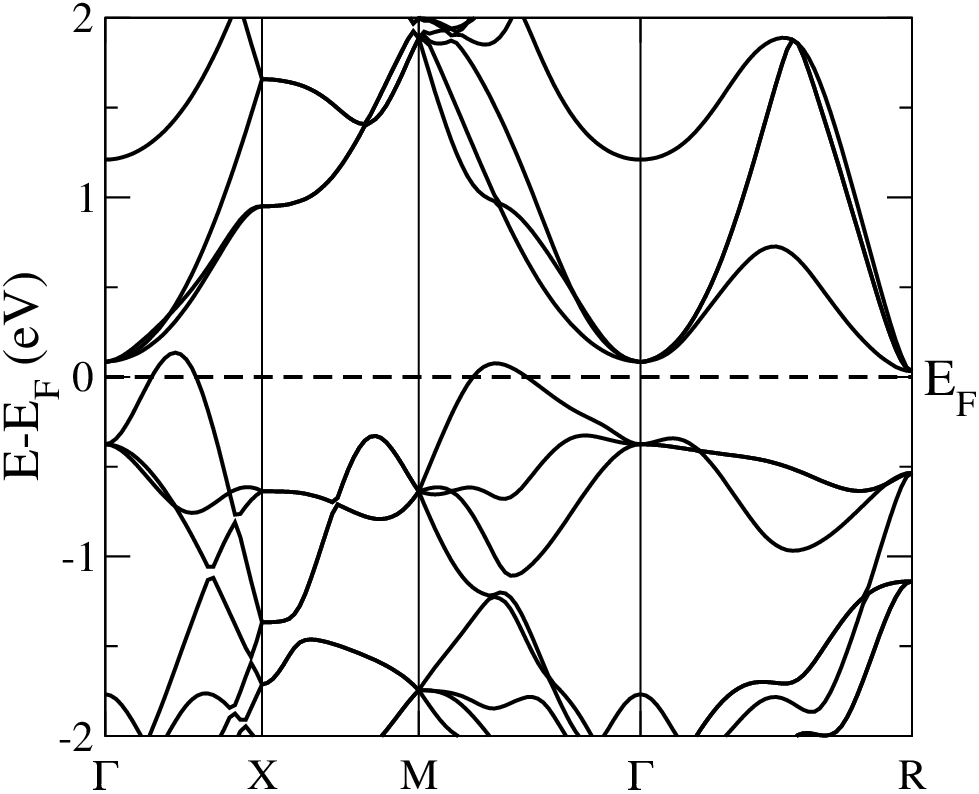}
\caption{Band structure of doped Pt$_{0.99}$Ir$_{0.01}$Sb$_2$.}
\end{figure}

\section{computational method}
We calculate the band structure of PtSb$_2$ using density function theory as implemented in the WIEN2k package \cite{wien2k}. The popular generalized gradient approximation \cite{GGA}is employed to optimize the volume and the internal atomic coordinates. To simulate doping, we use the virtual crystal approximation \cite{VCA} and rigid band approach
 \cite{RBA1,RBA2}. This approximation is widely employed in calculations of transport properties of doped semiconductors and is accurate when the doping is not too large. After having reached 
self-consistency in the calculations, we employ the post-processing BoltzTraP code \cite{BoltzTraP} to calculate the thermopower. This tool has demonstrated quantitative accuracy in calculating
 the thermopower of metals and doped semiconductors \cite{Madson, Scheidemantel, Singh, Zhang}. We use 3000 \textbf{k}-points in the full Brillouin zone for calculating the electronic structure
 and a dense mesh of 3564 \textbf{k}-points in the irreducible Brillouin zone for the thermoelectric calculations. PtSb$_2$ crystallizes in a cubic pyrite structure with space group $Pa\bar{3}$ and 
therefore exhibite isotropic transport properties. Our optimized lattice constant is 6.47 \AA\ which is close to the experiment value of 6.44 \AA\ \cite{Brese}.  

\begin{figure}
\includegraphics[width=0.5\textwidth,clip]{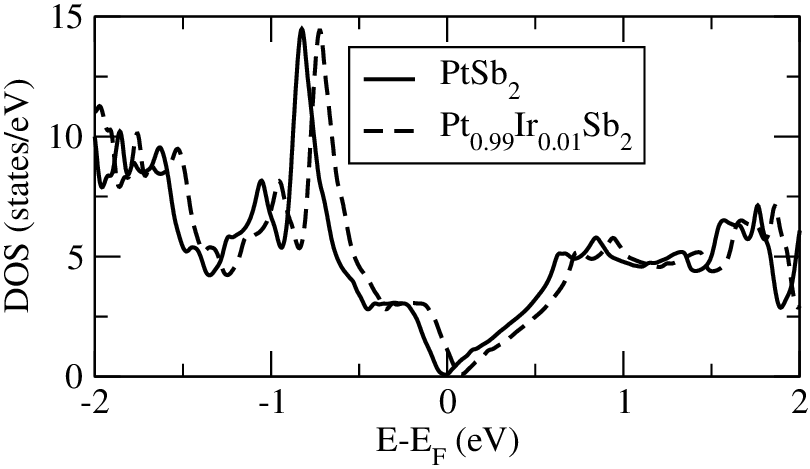}
\caption{Density of states of undoped and doped PtSb$_2$.}
\end{figure}
\section{Results and discussion}
The metallic states are reproduced by our band structure calculations for Pt$_{0.99}$Ir$_{0.01}$Sb$_2$ Fig. 1, while the undoped PtSb$_2$ is an insulator with experimental gap of 110 meV at T$\approx$10 K \cite{Reynolds}, already a small doping of 0.04 holes per unit cell resulting in a metallic state \cite{Nishikubo}. similarly, for a small $n$-doping also results in a metallic state \cite{Sondergaard}. The bands near the Fermi energy ($E_F$) mainly are due to the Sb 5$p$ orbitals, with some admixtures of the Pt 5$d$ orbitals. The band crosses the $E_F$ originates completely from the Sb 5$p$ orbitals across the symmetry lines $\Gamma$-X-M-$\Gamma$. Mori $et$ $al.$ \cite{Mori} have shown that bands at $E_F$ are corrugated bands and not the pudding mold type bands which establish the high Seebeck coefficients in Na$_x$CoO$_2$ \cite{pudding} and K$_x$RhO$_2$ \cite{krho}. Calculated densities of state (DOS) of PtSb$_2$ and Pt$_{0.99}$Ir$_{0.01}$Sb$_2$ are shown in Fig. 2. It is clearly visible that doping does not change the shape of the DOS but only the position of $E_F$.

\begin{figure}
\includegraphics[width=0.6\textwidth,clip]{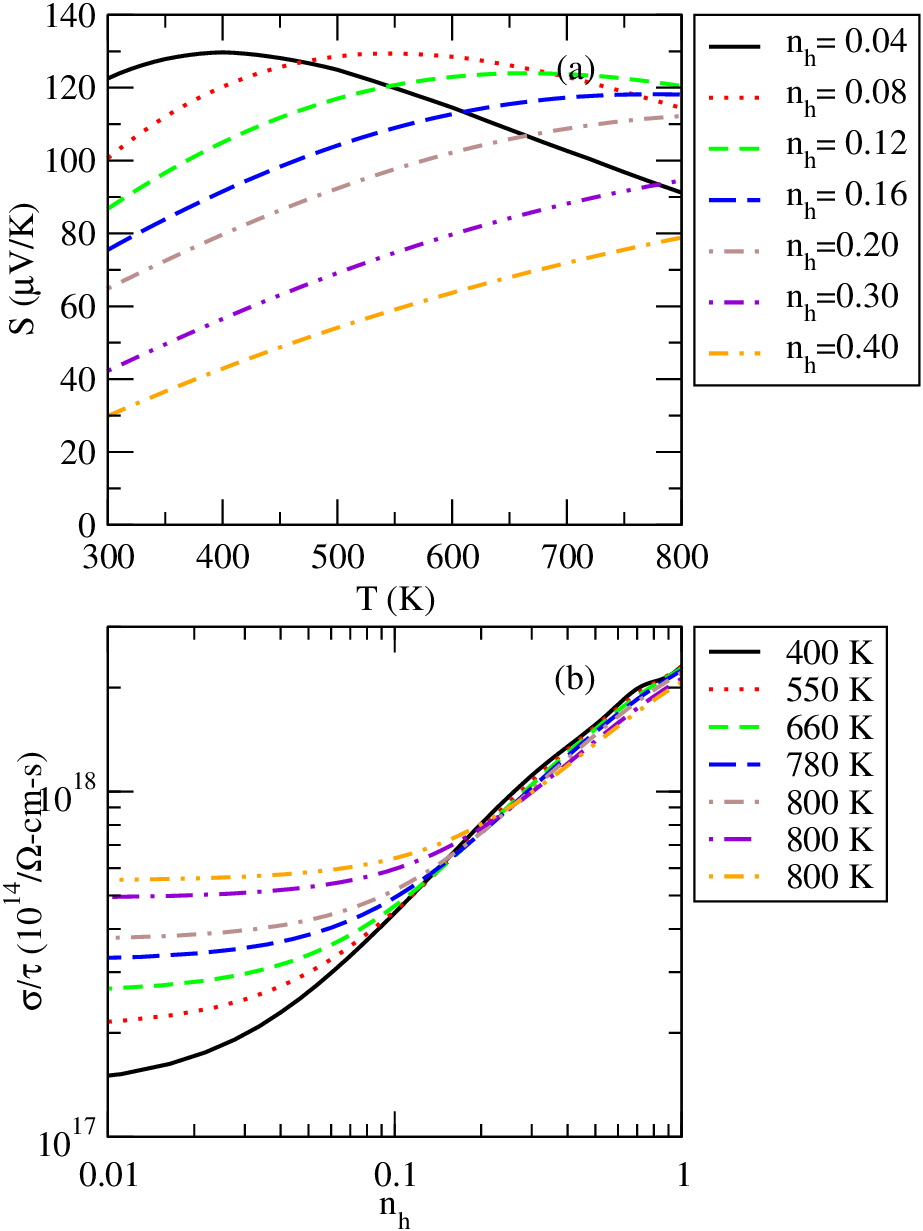}
\caption{Calculated $S$ and $\sigma/\tau$ of hole doped PtSb$_2$. The number of holes per unit cell is denoted by $n_h$.}
\end{figure}
\begin{figure}
\includegraphics[width=0.6\textwidth,clip]{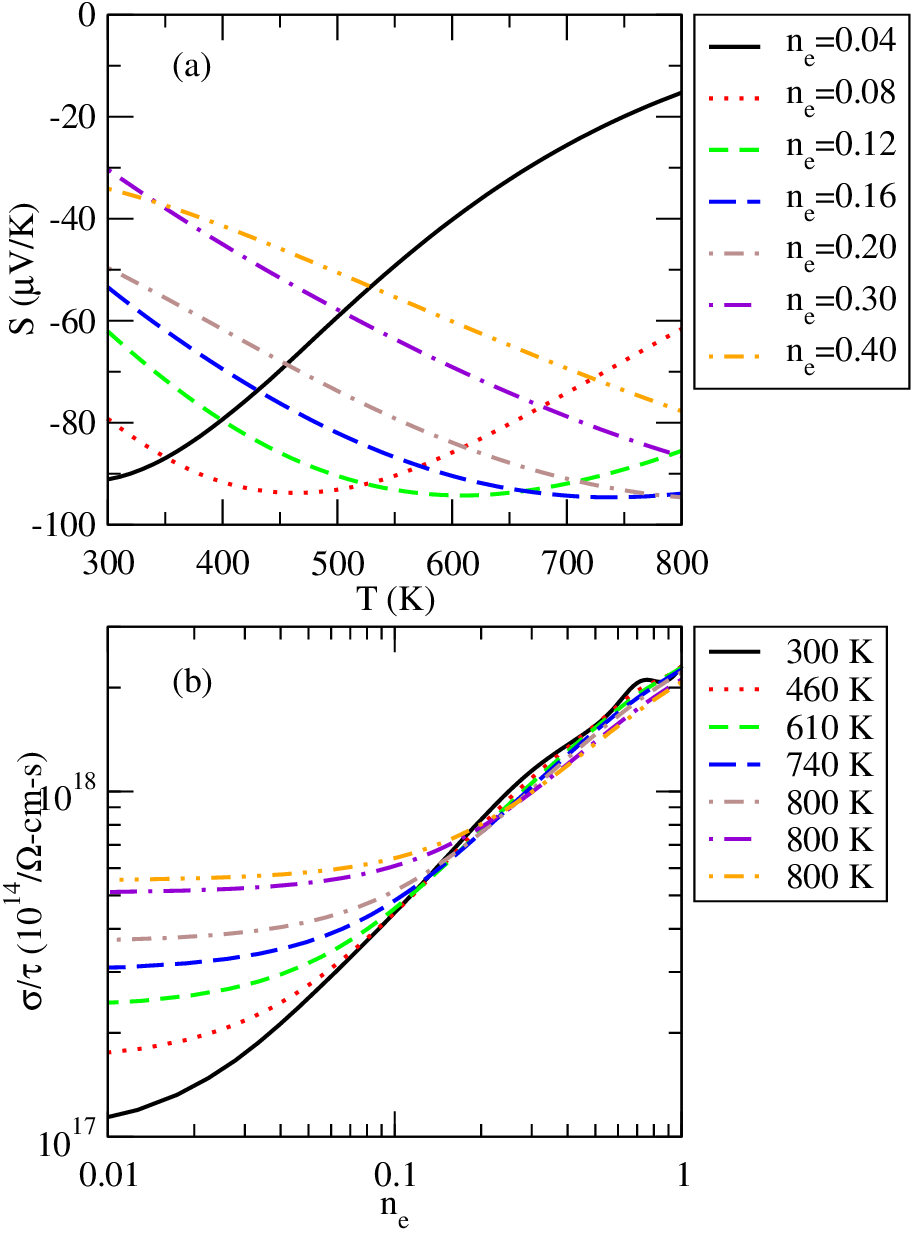}
\caption{Calculated $S$ and $\sigma/\tau$ of electron doped PtSb$_2$. The number of electrons per unit cell is denoted by $n_e$.}
\end{figure}

The calculated Seebeck coefficient of PtSb$_2$ in a doping range from 0.04 to 0.40 electrons/holes per unit cell for the temperature range from 300 to 800 K is plotted in Fig. 3(a)/4(a). 
Note that a doping of 0.04 electrons/holes per unit cell is equivalent to a carrier concentration of 1.5$\times$10$^{20}$ cm$^{-3}$ (as obtained experimentally for Pt$_{0.99}$Ir$_{0.01}$Sb$_2$ \cite{Nishikubo}).
 We find that $S$ shows a maximum of above 100 $\mu$V/K for each level of doping except for doping 0.3 and 0.4 electrons/holes per unit cell, but at different temperature. This maximum is due to the fact that the conduction band close to $E_F$ gives a 
negative contribution to the Seebeck coefficient at different temperatures and doping levels.

For $n_{h}=0.04$ holes per unit cell our calculation gives at room temperature a value of $S=122.5$ $\mu$V/K, which is in agreement
 with the experiment (100 $\mu$V/K) \cite{Nishikubo}. Our maximum $S$ value is 129.7 $\mu$V/K, while the experimental value is 112 $\mu$V/K at 400 K. 
 At $n_{h}=0.04$ doping, $S$ remains high upto 450 K, decrease thereafter. Overall Fig. 3(a) shows that with increasing hole doping the Seebeck coefficient decreases at room temperature, which is also consistent with the experiment.
 Importantly, the maximum $S$ value can be obtained at different temperatures by controlling the carrier concentration, which helps to achieve an optimal performance of the thermoelectric device under different conditions. In Fig. 3(b) we examine the effect of hole doping on the electrical conductivity $\sigma/\tau$ at room temperature and of the temperatures for which we have obtained the highest $S$ for each doping. We find variations with the temperature in the case of low doping, with the highest value at 800 K.
 For $n_{h}\gtrsim{0.20}$ the value of $\sigma/\tau$ increases rapidly and becomes virtually identical for the considered temperatures. Due to a sharp increase in $\sigma/\tau$, $S$ is reduced for high doping.
\begin{figure}
\includegraphics[width=0.9\textwidth,clip]{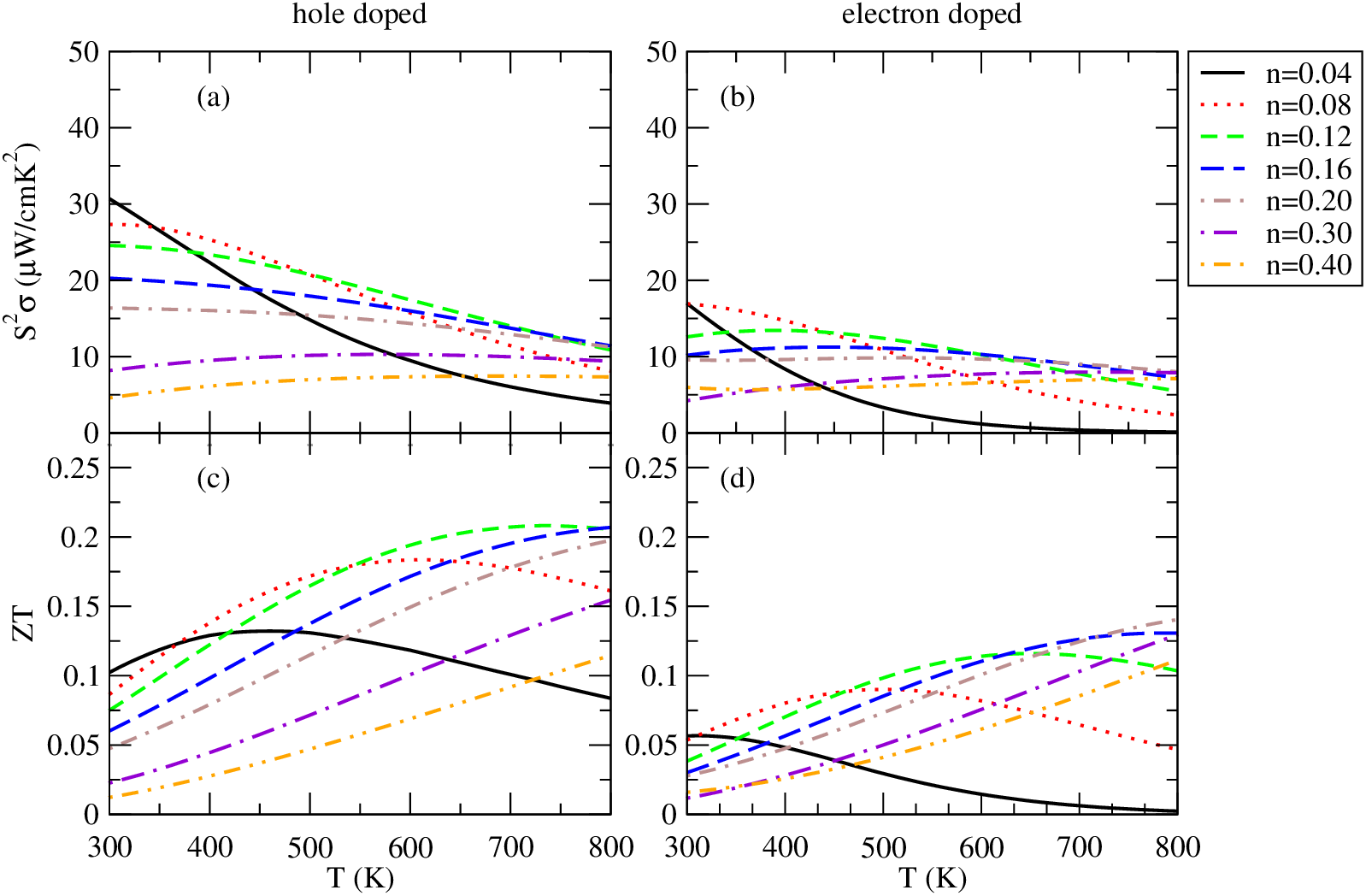}
\caption{Calculated $S^2\sigma$ and $ZT$ for electron and hole doped PtSb$_2$ as functions of the temperature for all doping level.}
\end{figure}

As it is experimentally known that both types of doping are possible in PtSb$_2$, we also consider electron doping, which has not been addressed experimentally.
 Nishikubo $et$ $al.$ \cite{Nishikubo} have shown that the substitution of Sn at the Sb site results in a small power factor.
This suggests that substitution at the Pt site is more favorable for the transport properties. Hence, we study substitution of Au at the Pt site, see Fig. 4(a,b). 
At the room temperature for $n_{e}$ 0.04, 0.12, and 0.40, $S$ is $-$91 $\mu$V/K, $-$62 $\mu$V/K, and  $-$34 $\mu$V/K, respectively. 
This decreasing trend of $S$ with increasing doping at room temperature is similar to hole doping. The maximum $S$ is always found around $-$94 $\mu$V/K except for $n_{e}$ 0.3 and 0.4 under the consider temperature range and shifts more and more to higher temperatures with increasing doping. 
The variation of $\sigma/\tau$  with the electron concentration with the temperature is similar to the case of hole doping.

It is fruitful to compare electron and hole doping in PtSb$_2$ for possible applications in thermoelectric generators with same host material as electron and hole conduction.
 In Fig. 5(a,b,c,d) we present the $S^{2}\sigma$ and $ZT$ of doped PtSb$_2$, with doping range from 0.04-0.40 electrons/holes per unit cell and temperature between 300 K and 800 K. For more realistic contribution of conductivity, we vary mobility with respect to carrier concentration and temperature. Whereas the electronic contribution $\kappa_{el}$ deduce from Wiedemann-Franz relation and phonon contribution take from experiment and vary as $\kappa_{ph}\alpha$ $T^{-1}$. Details of methodology can be found in Ref.\cite{pbse}.
In Fig. 5(a), the calculated room temperature $S^{2}\sigma$ for $n_{h}=0.40$ is 31 $\mu$W/cmK$^2$ 
in excellent agreement with the experimental value of 35 $\mu$W/cmK$^2$. For all other doping, $S^{2}\sigma$ values follow the trend of experiment at room temperature, decreases with temperature. In case of electron doped PtSb$_2$, $S^{2}\sigma$ is reaches upto 17 $\mu$W/cmK$^2$ for lower doping which half of the hole doped PtSb$_2$, while at higher doping electron and hole doped PtSb$_2$ have the same magnitude of $S^{2}\sigma$ at room temperature. Importantly, $S^{2}\sigma$ behavior changes inversely with temperature for doping level 0.04 electron/hole per unit cell with respect to other doping levels. While the experiment report the power factor only for x = 0.01 above 300 K. Room temperature $ZT$ for doping of 0.04 holes per unit cell is above 0.1 which is in good agreement with the experimental value of 0.12. Hence, we believe our calculated results at higher temperature will also be correct. Doping between 0.12$-$0.20 electron/hole per unit cell in PtSb$_2$, a large $S^{2}\sigma$ 
at 800 K results in a high ZT of around 0.10 and 0.20, respectively, which is more than double than the room temperature values of $ZT$. In order to further increase $ZT$ by reducing thermal conductivity, a nano structure 
PtSb$_2$ could be a better choice for thermoelectric devices.
   
Having the same host material with different types of doping joint in a device a high efficiency can be expected because of low losses at the thermal contact. 
This implies that PtSb$_2$ is a good candidate for thermoelectrics. We estimate the thermoelectric device efficiency 
of doped PtSb$_2$ at $T_{hot}=800$ K, $T_{cold}=300$ K ($T_{avg}=400$ K) for $n$ 0.08, 0.12 and 0.16, which shows high $ZT$ at $T_{avg}$. For $T_{avg}=400$ K we obtain thermoelectric generator efficiency of 1.07$\%$, 0.82$\%$ and 0.62$\%$ for doping 0.08, 0.12 and 0.16, respectively. This efficiency could be increased by As alloying at Sb site, by reducing the thermal conductivity and by increasing Seebeck (by increasing the band gap).   

\section{Conclusion}
To conclude, we have studied the transport properties of electron and hole doped PtSb$_2$ over a wide doping range,  using first principles calculations. 
A doping of 0.04 electrons/holes per unit cell gives a high power factor at room temperature, which decreases further with increasing temperature. 
Our results show that doping between 0.12$-$0.20 electrons/holes per unit cell, the $ZT$ are more than double achieved at 800 K due to a high electrical conductivity. 
Experiments should be performed at this doping level and temperature range for confirmation. We have demonstrated that the highest Seebeck coefficient is obtained at different 
temperatures for different doping levels. This allows to optimize thermoelectric devices operated at different temperatures by tuning the carrier 
concentration. We also calculate thermoelectric generator efficiency for certain doping levels. Moreover, it will be desirable to study in more detail about nano structure doped PtSb$_2$ to further reduce thermal conductivity.

\end{document}